\documentclass[aps,twoside,twocolumn,fleqn,floatfix,showpacs,superscriptaddress]{revtex4}
\usepackage{amssymb}
\usepackage{bm}
\usepackage{graphicx}

\def\buch{Institute for Nuclear Physics and Engineering, Bucharest, Romania}
\def\buda{KFKI Research Institute for Particle and Nuclear Physics, Budapest, Hungary}
\def\cler{Laboratoire de Physique Corpusculaire, IN2P3/CNRS,
 and Universit\'{e} Blaise Pascal, Clermont-Ferrand, France}
\def\darm{Gesellschaft f\"{u}r Schwerionenforschung, Darmstadt, Germany}
\def\dres{IKH, Forschungszentrum Rossendorf, Dresden, Germany}
\def\heid{Physikalisches Institut der Universit{\"a}t Heidelberg, Heidelberg, Germany}
\def\mosc{Institute for Theoretical and Experimental Physics, Moscow, Russia}
\def\kurc{Kurchatov Institute, Moscow, Russia}
\def\seou{Korea University, Seoul, Korea}
\def\stra{Institut de Recherches Subatomiques and
 Universit\'{e} Louis Pasteur, Strasbourg, France}
\def\wars{Institute of Experimental Physics, Warsaw University, Warsaw,Poland}
\def\zagr{Ru{d\llap{\raise 1.22ex\hbox
    {\vrule height 0.09ex width 0.2em}}\rlap{\raise 1.22ex\hbox
    {\vrule height 0.09ex width 0.06em}}}er
    Bo\v{s}kovi\'{c} Institute, Zagreb, Croatia}

\begin{document}

\title[]{Charged pion production in $^{96}_{44}$Ru + $^{96}_{44}$Ru 
collisions at 400$A$ and 1528$A$ MeV}

\author{B. Hong} \email{bhong@korea.ac.kr} \affiliation{\seou}
\author{Y.J. Kim} \affiliation{\seou} \affiliation{\darm}
\author{N. Herrmann} \affiliation{\heid} 
\author{M.R. Stockmeier} \affiliation{\heid}
\author{A. Andronic} \affiliation{\darm}
\author{V. Barret} \affiliation{\cler}
\author{Z. Basrak} \affiliation{\zagr}
\author{N. Bastid} \affiliation{\cler}
\author{M.L. Benabderrahmane} \affiliation{\heid}
\author{R. \v{C}aplar} \affiliation{\zagr} 
\author{P. Crochet} \affiliation{\cler} 
\author{P. Dupieux} \affiliation{\cler}
\author{M. D\v{z}elalija} \affiliation{\zagr} 
\author{Z. Fodor} \affiliation{\buda}
\author{A. Gobbi} \affiliation{\darm} 
\author{Y. Grishkin} \affiliation{\mosc} 
\author{O.N. Hartmann} \affiliation{\darm} 
\author{K.D. Hildenbrand} \affiliation{\darm} 
\author{J. Kecskemeti} \affiliation{\buda} 
\author{M. Kirejczyk} \affiliation{\wars} \affiliation{\darm}
\author{P. Koczon} \affiliation{\darm} 
\author{M. Korolija} \affiliation{\zagr} 
\author{R. Kotte} \affiliation{\dres} 
\author{T. Kress} \affiliation{\darm}
\author{A. Lebedev} \affiliation{\mosc} 
\author{Y. Leifels} \affiliation{\darm}
\author{X. Lopez} \affiliation{\cler}
\author{A. Mangiarotti} \affiliation{\heid} 
\author{M. Merschmeyer} \affiliation{\heid} 
\author{W. Neubert} \affiliation{\dres}
\author{D. Pelte} \affiliation{\heid}
\author{M. Petrovici} \affiliation{\buch} 
\author{F. Rami} \affiliation{\stra}
\author{W. Reisdorf} \affiliation{\darm}
\author{A. Sch\"{u}ttauf} \affiliation{\darm}
\author{Z. Seres} \affiliation{\buda}
\author{B. Sikora} \affiliation{\wars} 
\author{K.S. Sim} \affiliation{\seou}
\author{V. Simion} \affiliation{\buch}
\author{K. Siwek-Wilczy\'{n}ska} \affiliation{\wars}
\author{V. Smolyankin} \affiliation{\mosc} 
\author{G. Stoicea} \affiliation{\buch} 
\author{Z. Tyminski} \affiliation{\wars} \affiliation{\darm}
\author{P. Wagner} \affiliation{\stra} 
\author{K. Wi\'{s}niewski} \affiliation{\wars} 
\author{D. Wohlfarth} \affiliation{\dres}
\author{Z.G. Xiao} \affiliation{\darm}
\author{I. Yushmanov} \affiliation{\kurc}
\author{A. Zhilin} \affiliation{\mosc}

\collaboration{FOPI Collaboration}
\noaffiliation

\date[]{Received October 12, 2004}

\begin{abstract}
We present transverse momentum and rapidity spectra of 
charged pions in central Ru + Ru collisions at 400$A$ 
and 1528$A$ MeV. 
The data exhibit enhanced production 
at low transverse momenta compared to the expectations 
from the thermal model that includes the decay of 
$\Delta(1232)$-resonances and thermal pions.
Modification of the $\Delta$-spectral function and 
the Coulomb interaction are necessary to describe 
the detailed shape of the transverse momentum spectra.
Within the framework of the thermal model, 
the freeze-out radii of pions are similar at both beam energies. 
The IQMD model reproduces the shapes of the transverse 
momentum and rapidity spectra of pions, but the predicted 
absolute yields are larger than in the measurements, 
especially at lower beam energy.
\end{abstract}


\maketitle

\section{introduction}\label{intro}

Collision of relativistic heavy ions is a unique method to 
produce a large volume of excited nuclear matter in 
the laboratory at present. 
At incident beam energies near 1 GeV per nucleon
nuclear matter can reach its density about two to three 
times higher than normal nuclear matter density 
at the temperature below 100 MeV \cite{rst1,hst1}. 
Major motivations for such studies aim to determine 
the nuclear equation of state (EoS) and to study the 
basic properties of quantum chromodynamics (QCD), 
the theory of strong interactions. 
These informations are important not only in their own virtue 
but also for understanding the behavior of astrophysical objects, 
such as neutron stars and supernovae \cite{gbr1}.

Experiments of this kind started with various 
ion beams at the BEVALAC in 70's \cite{bev1,bev2}.
Since the early 90's the heavy ion synchrotron SIS 
at GSI-Darmstadt, Germany, took over a leading role 
in relativistic heavy ion collisions in the energy range 
up to 2$A$ GeV. 
A scenario of several complementary experimental 
setups at SIS has allowed to perform thorough 
investigations of the EoS and in-medium 
properties of hadrons \cite{gsi1,gsi2,fpi1,fpi2,fopi-jkps}.

Several observables, which are accessible in experiments, 
have been proposed as sensitive probes to characterize 
the properties of hot and dense nuclear matter. 
Among the prominent candidates are the collective 
flow \cite{wre1,jol1} and the particle production \cite{psen1}.
Especially the production of pions has been suggested 
rather early \cite{stoe1}, as their yield can be connected 
to the temperature of the fireball through the nucleon resonances. 
However, later detailed theoretical investigations revealed that 
pions might not be so sensitive to the EoS \cite{bert1,kruse1}.
Nevertheless the pion production in heavy ion collisions 
has attracted continuous attention because it is the most 
important inelastic channel in nuclear collisions \cite{dio,e895_pi}.
The description of this process is necessary to 
understand the whole dynamic evolution of the fireball 
from the early stage to the freeze-out.

In this paper we present results on the pion production 
in Ru + Ru collisions at 400$A$ and 1528$A$ MeV. 
In the analysis we want to test in particular 
the modification of the $\Delta$-spectral function 
in heavy ion collisions \cite{wein1,wein2} 
and the Coulomb interaction \cite{wagn1} 
within the framework of the thermal model.
Previously, similar ideas have been applied to the pion 
spectra in Au + Au collisions at 1$A$ GeV by the KaoS 
collaboration at SIS/GSI \cite{wein1,wein2,wagn1,hong4}
and at 10.8$A$ GeV by the E877 collaboration at 
AGS/BNL \cite{e877-1}. 
The experimental data are also compared to microscopic 
transport model calculations, namely the Isospin 
Quantum Molecular Dynamics (IQMD) \cite{jai1,hart1,bass1}. 
Finding the common features in the interpretation of 
experimental data by using these two completely 
different approaches may shed some light on the process 
of particle production in relativistic nuclear collisions.

In Sec.~\ref{expt}, the experimental setup and the method 
for the selection of collision centrality are described. 
The main experimental results on the pion spectra as 
functions of transverse momentum and rapidity are 
presented in Sec.~\ref{spectra}. 
In Secs.~\ref{thermal} and~\ref{iqmd} we compare 
the experimental data with the thermal model and 
the IQMD calculations, respectively.
Finally, conclusions follow in Sec.~\ref{conc}.

\section{experiment}\label{expt}

Collisions of $^{96}_{44}{\rm Ru}$ nuclei with 
a $^{96}_{44}{\rm Ru}$ target of 380 mg/cm$^{2}$ thickness 
were studied at 400$A$ and 1528$A$ MeV with the FOPI detector 
at SIS. The beam intensities were typically on the order of 
3$\times$10$^4$ ions/sec. A similar amount of events for 
`central' and `medium central' conditions was accumulated 
(some under `minimum bias' conditions).
The central trigger required a high multiplicity in the forward 
Plastic Wall (PLAWA), which covers the laboratory polar angles
$\theta_{lab} = 7^{\circ} - 27^{\circ}$ with full azimuthal 
symmetry, corresponding to about 15 \% of the total geometric 
cross section $\sigma_{geom}$. The FOPI detector system 
is described in detail elsewhere \cite{aga1,jri1}.

For the tracking of charged particles we use the 
central drift chamber (CDC) which is placed inside a uniform 
solenoidal magnetic field of 0.6 T. 
The CDC covers $\theta_{lab} = 32^{\circ} - 140^{\circ}$. 
Pions, protons, deuterons and heavier particles are identified 
by using the correlation of the specific energy loss ($dE/dx$) 
and the magnetic rigidity (the laboratory momentum $p$ divided 
by charge) determined by the curvature of the tracks. 
Details of the detector resolution and performance can be 
found in Refs. \cite{aga1,jri1,hong1,best1}.

\begin{figure}[t!]
\begin{center}
\includegraphics[width=8.5cm,height=4.7cm]{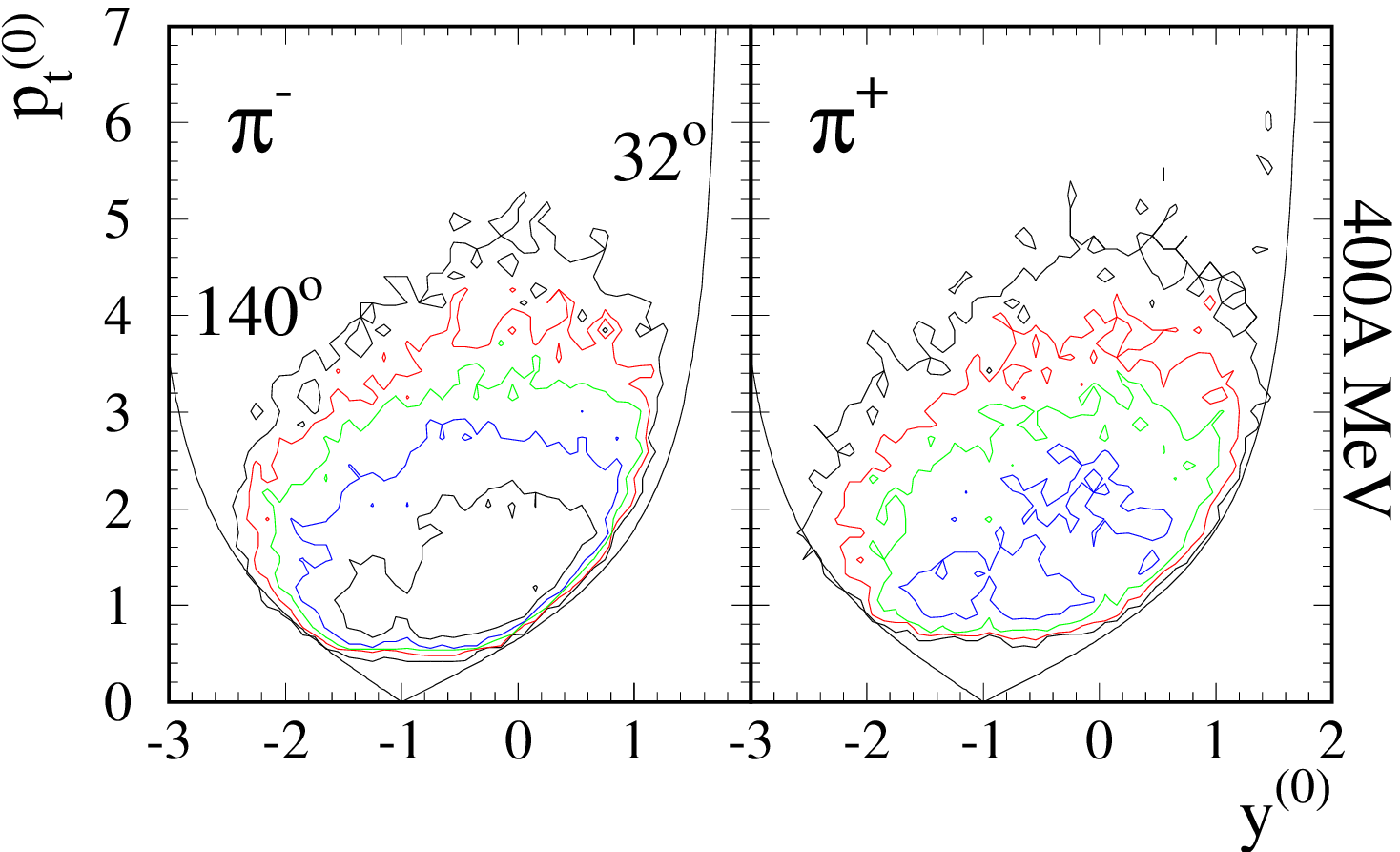}
\includegraphics[width=8.5cm,height=4.7cm]{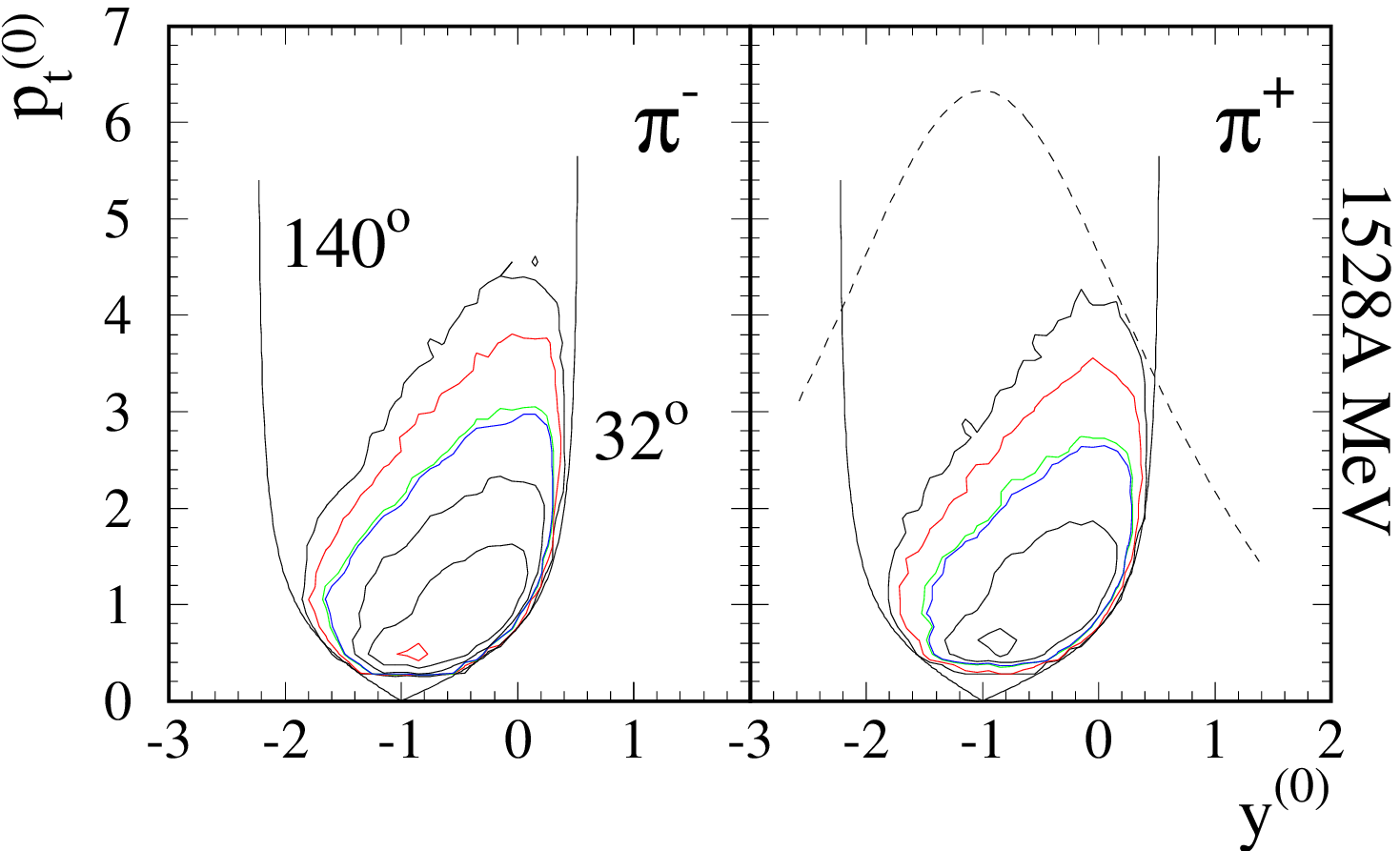}
\end{center}
\caption[]{
Measured raw yields in the plane of normalized transverse momentum vs. 
normalized rapidity of $\pi^{-}$ (left) and $\pi^{+}$ (right) 
in Ru + Ru collisions at 400$A$ (top) and 1528$A$ MeV (bottom).
In all cases the most central 10 \% of $\sigma_{geom}$ 
have been selected. Two solid lines in each panel show the 
geometrical limits of the CDC. The dashed line in the lower 
right panel represents the upper limit of the pion laboratory 
momentum (800 MeV) in order to separate $\pi^{+}$'s 
from protons. Each successive contour line represents 
a relative factor of two in yields.}
\label{Fig:fpi_acc}
\end{figure}

To illustrate the measured phase space we show 
the CDC acceptance of charged pions at both beam 
energies in Fig.~\ref{Fig:fpi_acc}. 
In this figure, $p_{t}^{(0)}$ represents 
the normalized transverse momentum calculated by 
$p_{t}/(m_{\pi}\beta_{cm}\gamma_{cm})$, 
where $p_{t}$ is the transverse momentum,
$m_{\pi}$ is the pion mass, 
and $\beta_{cm}$ and $\gamma_{cm}$ are 
velocity and Lorentz gamma factor of 
the center of mass (c.m.), respectively.
In addition, $y^{(0)}$ represents the normalized rapidity 
$y_{lab}/y_{cm} - 1$, where $y_{lab}$ and $y_{cm}$ are 
the pion rapidity in the laboratory frame and 
the c.m. rapidity of the collision system, respectively. 
As a result, the CDC covers more than 90 \% of 
the full solid angle when the symmetry of the colliding system 
around midrapidity is utilized.
For both beam energies a target absorption effect is 
visible at small $p_{t}$ values near the target 
rapidity region ($-1.2 \leq y^{(0)} \leq -0.8$).

The collision centrality of each event is determined by 
two methods depending on beam energy.
At 400$A$ MeV the variable $E_{rat}$ is used, defined as ratio 
of total transverse ($E_{\perp}$) to longitudinal kinetic energy 
($E_{\parallel}$) in the center of mass:
\begin{equation}
E_{rat}=\sum_{i}E_{\perp,i}/\sum_{i}E_{\parallel,i},
\label{erat}
\end{equation}
where $i$ runs over all detected charged particles in the CDC 
and the PLAWA. Previously, it has been demonstrated that 
$E_{rat}$ is a suitable variable for event centrality, especially 
in central collisions at beam energies $\leq 400A$ MeV \cite{wre2}. 
At 1528$A$ MeV the total multiplicity seen 
in the CDC and the PLAWA is used. 
In this paper we select events only for the upper most 380 mb, which 
corresponds to $\sim$ 10 \% of the total geometric cross section. 
Total numbers of analyzed events under these cuts are
approximately 80000 at 400$A$ MeV and 240000 at 1528$A$ MeV. 
We will adopt the natural units $\hbar = c = 1$ in the following.

\section{results} \label{spectra}

\begin{figure}[t!]
\begin{center}
\includegraphics[width=8.8cm]{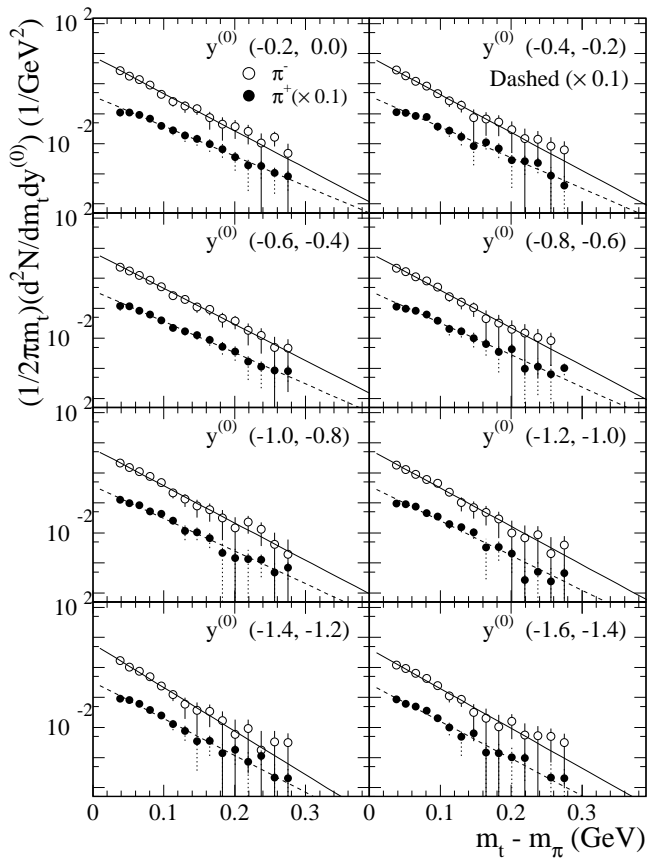}
\end{center}
\caption[]{
Invariant spectra of $\pi^{-}$ (open circles) and $\pi^{+}$ 
(solid circles) as a function of $m_{t}-m_{\pi}$ in Ru + Ru collisions 
at 400$A$ MeV for the most central 10 \% of $\sigma_{geom}$. 
The error bars represent the statistical and $p_{t}$-dependent 
systematic errors. Solid and dashed lines are exponential fit functions 
($p_{t} \geq$ 120 MeV) to the invariant spectra of $\pi^{-}$ and 
$\pi^{+}$, respectively. The $\pi^{+}$ data and fit functions are 
scaled by a factor of 0.1 for a clearer display.}
\label{Fig:efit4}
\end{figure}

\begin{figure}[t!]
\begin{center}
\includegraphics[width=8.8cm]{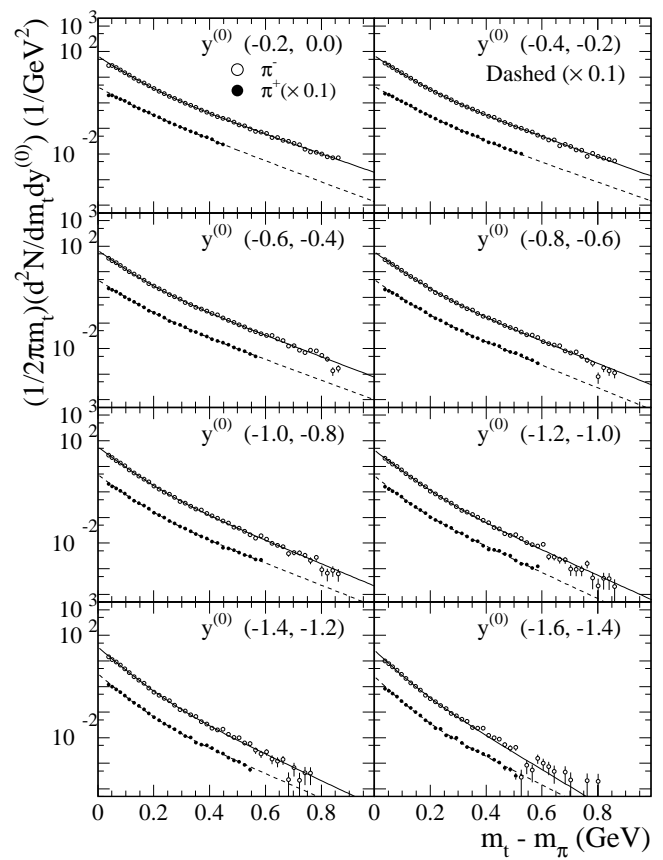}
\end{center}
\caption[]{
Same as Fig.~\ref{Fig:efit4}, but for 1528$A$ MeV.
The $p_{t}$-dependent systematic errors are negligible at this 
beam energy, so the error bars are dominated by statistics.
Solid and dashed lines are the sum of two exponential 
fit functions ($p_{t} \geq$ 120 MeV) to the invariant spectra of 
$\pi^{-}$ and $\pi^{+}$, respectively. The $\pi^{+}$ data and 
fit functions are scaled by a factor of 0.1 for a clearer display.}
\label{Fig:efit15}
\end{figure}

\begin{figure}[t!]
\begin{center}
\includegraphics[width=8.5cm]{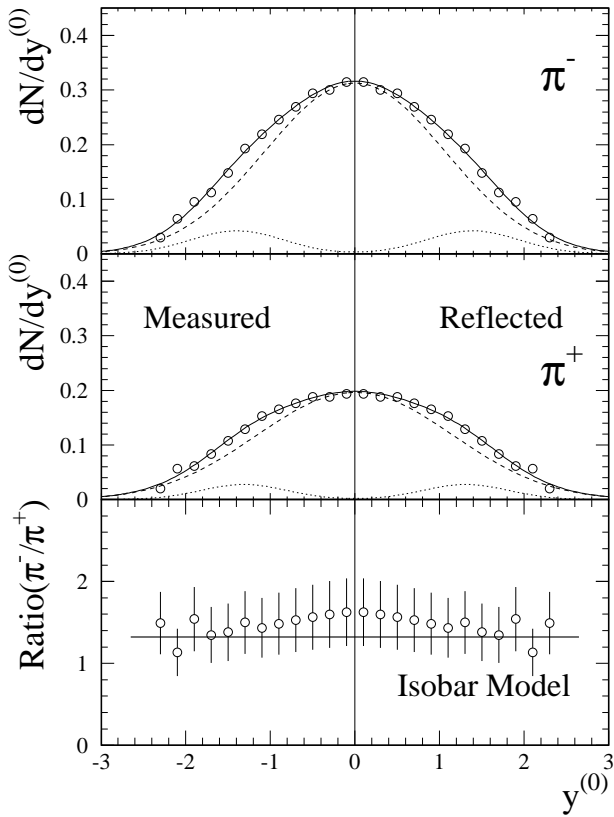}
\end{center}
\caption[]{
Rapidity distributions of $\pi^{-}$ (top) and 
$\pi^{+}$ (middle) in Ru + Ru collisions at 400$A$ MeV 
for the most central 10 \% of $\sigma_{geom}$. 
Dashed and dotted lines in the two upper panels are 
the calculations for an isotropic thermal source at the center of mass 
and the estimated target/projectile components, respectively. 
Solid lines represent the sums of two contributions. 
The bottom panel shows the ratio of the rapidity distributions of 
$\pi^{-}$ and $\pi^{+}$. The error bars in this panel consider 
both statistical and systematic errors, but the dominant contribution 
is given by the systematic errors, however. The solid horizontal line 
at 1.32 represents the estimated ratio of the isobar model.}
\label{Fig:dndy4}
\end{figure}

\begin{figure}[t!]
\begin{center}
\includegraphics[width=8.5cm]{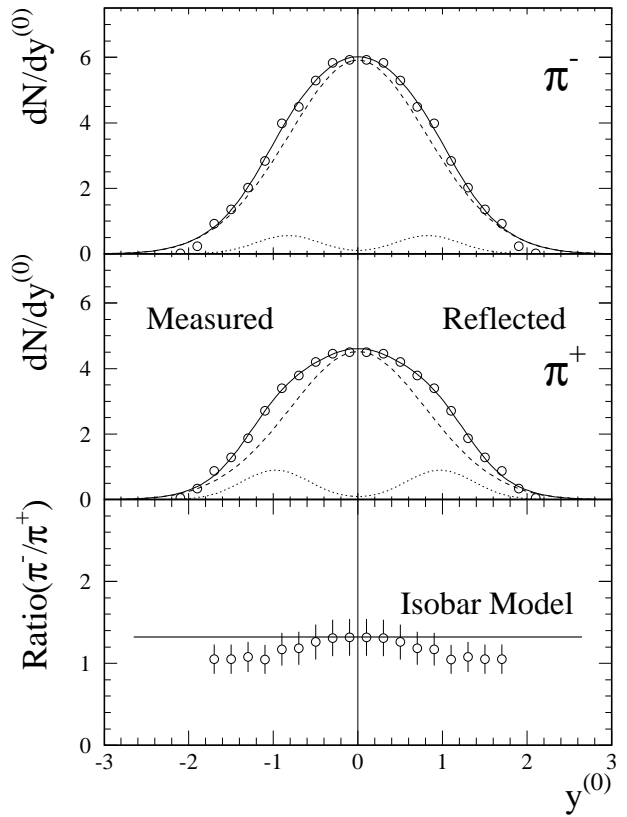}
\end{center}
\caption[]{
Same as Fig.~\ref{Fig:dndy4}, but for 1528$A$ MeV.}
\label{Fig:dndy15}
\end{figure}
 
\begin{table*}[t!]
\caption{Inverse slope parameter $T_{0}$ and 
the measured and the extrapolated numbers of $\pi^{\pm}$ 
per event at midrapidity ($-0.2 < y^{(0)} \leq 0.0$) 
for the most central 10 \% of $\sigma_{geom}$.
The statistical errors are negligible whereas 
the estimated systematic errors are 18 and 12 \% 
at 400$A$ and 1528$A$ MeV, respectively.
For the later comparison, the $p_{t}$-integrated results 
in the same rapidity interval from the IQMD calculations 
are also included. For IQMD, HM and SM represent 
a hard and a soft EoS, respectively, including 
the momentum dependence of the nucleon interaction.}
\begin{ruledtabular}
\begin{tabular}{ccccccc}
 & & $T_{0}$ (MeV) & Measured data & Extrapolated data & IQMD(HM) & IQMD(SM) \\
 & & & ($p_{t} \geq 120$ MeV) & & & \\ \hline
400$A$ MeV & $\pi^{-}$ & 35 & 0.11 & 0.32 & 0.46 & 0.45 \\
 & $\pi^{+}$ & 44 & 0.09 & 0.20  & 0.31 & 0.31 \\ \hline
1528$A$ MeV & $\pi^{-}$ & 57 (low)/113 (high)& 3.81 & 5.92 & 7.04 & 7.70 \\
 & $\pi^{+}$ & 62 (low)/107 (high)& 3.11 & 4.50 & 5.63 & 6.18 \\
\end{tabular}\label{Tab:midyield}
\end{ruledtabular}
\end{table*}

Figures~\ref{Fig:efit4} and \ref{Fig:efit15} show the experimental 
charged pion spectra in invariant form as a function of 
the transverse mass $m_{t}$ ($= \sqrt{p_{t}^{2} + m_{\pi}^{2}}$)
for several rapidity bins. 
The spectra are corrected for the CDC track reconstruction 
efficiencies which are determined by a GEANT based 
Monte-Carlo simulation \cite{geant} taking into account the 
detailed response of the FOPI detector; the IQMD model was used 
as event generator \cite{hart1}. 
The same method of event selection and particle identification was 
applied in the data analysis and in the simulation.
The track reconstruction efficiencies of CDC for $\pi^{\pm}$ 
are evaluated by taking the ratio of the GEANT output to input spectra. 
This procedure has been applied in the two-dimensional space 
of $p_{t}$ and $y^{(0)}$.
On average these reconstruction efficiencies are determined to 
about 62 \% and 80 \% for $\pi^{+}$ and $\pi^{-}$, 
respectively, at 400$A$ MeV.
These values become somewhat higher at 1528$A$ MeV, namely
$\sim$ 74 \% for $\pi^{+}$ and $\sim$ 89 \% for $\pi^{-}$.
Note that the overall tracking efficiency depends on 
various parameters such as the rapidity, $p_{t}$ and the track density.
In general, the tracking efficiency for the algorithm
used in this analysis is worse for smaller momentum. 
Figures~\ref{Fig:efit4} and ~\ref{Fig:efit15} show the efficiency 
corrected spectra. More detailed discussions of the systematic errors 
on the invariant and the rapidity spectra will be given later
in this section.

The invariant spectra of pions at 400$A$ MeV exhibit a one-slope 
structure in all rapidity bins (Fig.~\ref{Fig:efit4}) whereas clearly 
two slopes are present at 1528$A$ MeV (Fig.~\ref{Fig:efit15}).
Consequently, at 400$A$ MeV, the invariant spectra for each rapidity 
slice are fitted by one exponential function as follows:
\begin{equation}
{{1} \over {2 \pi m_{t}}} 
{{d^{2} N} \over {dm_{t} dy^{(0)}}} 
= C(y^{(0)}) \cdot \exp [-{{(m_{t}-m_{\pi})} \over {T_{0}(y^{(0)})}}],
\label{efit}
\end{equation}
where $C$ and $T_{0}$ are the rapidity dependent normalization 
constant and the inverse slope parameter, respectively. 
At 1528$A$ MeV, a sum of two exponential functions is required 
to describe spectra. Solid and dashed lines in Figs.~\ref{Fig:efit4} 
and \ref{Fig:efit15} represent the corresponding fit functions 
to the $\pi^{-}$ and $\pi^{+}$ invariant spectra, respectively, 
in the range $p_{t} \geq$ 120 MeV.
Note that the one- and two-slope fit functions are 
the simplest phenomenological description of the pion data.
The fitted $T_{0}$ values at midrapidity are listed 
in Table~\ref{Tab:midyield}.
Since the parametrization by Eq.~(\ref{efit}) describes 
the pion spectra reasonably well over all measured $m_{t} - m_{\pi}$,
they are used to extract the rapidity distributions. 
In details, we integrate the fitted exponential functions from 
$p_{t} =$ 0 to $\infty$, hence also accounting for the missing 
$p_{t}$ region in the CDC acceptance (see Fig.~\ref{Fig:fpi_acc}).
The extrapolation to the missing low $p_{t}$ region is rather 
significant as shown in Table~\ref{Tab:midyield} at midrapidity.

The fully extrapolated rapidity distributions are shown in the two 
top panels of Fig.~\ref{Fig:dndy4} for 400$A$ MeV and 
Fig.~\ref{Fig:dndy15} for 1528$A$ MeV, where the forward c.m. 
spectra are reflected around midrapidity by using the mass 
symmetry of the collision system. The dashed lines 
in Figs.~\ref{Fig:dndy4} and \ref{Fig:dndy15} are 
the simplest thermal model predictions for 
an isotropic thermal source at the c.m. \cite{schne1}:
\begin{eqnarray}
{{dN_{th}} \over {dy^{(0)}}} & \propto & 
T^{3}~\biggl( {{m_{\pi}^{2}} \over {T^{2}}} + 
{{m_{\pi}} \over {T}} {{2} \over {\cosh y_{C}}} +
{{2} \over {\cosh^{2} y_{C}}}\biggr) \nonumber \\
 & & \times \exp \biggl({{-m_{\pi}\cosh y_{C}} \over {T}}\biggr),
\label{dndy_therm}
\end{eqnarray}
with $y_{C} = y_{lab} - y_{cm}$ and $T = T_{0}$ at midrapidity
(higher $T_{0}$ component at 1528$A$ MeV). 
Here, $T$ is 35 (44) MeV at 400$A$ MeV and 113 (107) MeV 
at 1528$A$ MeV for $\pi^{-}$ ($\pi^{+}$). 
In Figs.~\ref{Fig:dndy4} and \ref{Fig:dndy15} 
the rapidity distributions from the isotropic thermal source 
are normalized at midrapidity. 
Obviously, for both beam energies the experimental rapidity 
distributions of $\pi^{\pm}$ are wider than the isotropic 
thermal source sitting at midrapidity. 
Also represented by dotted lines are the estimated 
target/projectile components which are merely Gaussian fits to 
the differences between data and the dashed lines. 
The sums of the two contributions are shown by solid lines 
in Figs.~\ref{Fig:dndy4} and \ref{Fig:dndy15}.

Our strategy to integrate the exponential fits of the invariant spectra 
for the final $dN/dy^{(0)}$ distributions has been justified by 
the GEANT based Monte-Carlo simulation \cite{geant}; 
the input rapidity spectra agree nicely with the $p_{t}$-integrated 
$dN/dy^{(0)}$ distributions obtained by the exponential fits to 
the efficiency corrected GEANT output invariant spectra. 
Nevertheless the most significant source of systematic errors 
for the $dN / dy^{(0)}$ distribution is the uncertainty in the estimation 
of the CDC efficiency which relies on the tracking strategy.
Such a systematic error is significant at 400$A$ MeV,
but it becomes much smaller at 1528$A$ MeV, however.
Furthermore, this uncertainty depends on $p_{t}$. 
It is negligible in the low $p_{t}$ region ($<$ 200 MeV), 
but increases with transverse momentum, reaching about 50 \% 
at $p_{t} =$ 300 MeV in case of the 400$A$ MeV beam energy
(the error bars in Figs.~\ref{Fig:efit4} and \ref{Fig:efit15}
reflect the systematic as well as statistical errors).
As a result, the systematic error in the $p_{t}$-integrated 
$dN/dy^{(0)}$ distributions which is caused by 
the uncertainty of the CDC track reconstruction is estimated 
to about 15 \% at 400$A$ MeV and 6 \% at 1528$A$ MeV.
Other sources of systematic errors which are similar 
at both beam energies can be summarized as follows.
The systematic error in the particle identification (2 \%) 
is determined by changing various selection criteria 
for good event and track samples.
In addition, different fitting ranges in $p_{t}$ also 
cause maximal 2 \% uncertainty in $dN/dy^{(0)}$.
The error due to the extrapolation procedure over 
the complete $p_{t}$-range is estimated to be about 10 \%. 
For this estimation, we first use various fit functions, 
e.g., an exponential fit in the invariant and 
the Boltzmann representations. 
Furthermore, we also use the IQMD model; 
the difference between the extrapolated yields 
by the fit functions to the data and the pion yields 
by IQMD for $p_{t} <$ 120 MeV are less than 2 \% 
at both beam energies (note that, for this test, 
the IQMD spectra should be normalized to the measured 
pion spectra for $p_{t} >$ 120 MeV because 
the absolute yields by IQMD are always larger 
than the data as will be clear in Sec.~\ref{iqmd}). 
Employing the IQMD model for the extrapolation to 
lower $p_{t}$ can be supported by confirming that 
the yield ratios of $\pi^{-}$ to $\pi^{+}$ by IQMD 
(independent of the EoS) agree with the estimations 
by the fit function to the data within 5 \% 
for $p_{t} <$ 120 MeV at both beam energies. 
Finally, assuming that the sources of various systematic errors 
are incoherent, we calculate overall 
systematic errors for the $dN / dy^{(0)}$ of both charged pions 
of 18 \% and 12 \% at 400$A$ and 1528$A$ MeV, respectively,
by taking a quadratic sum of all contributions.
Only the statistical errors are shown for the spectra presented 
in this paper, unless explicitly noted differently. 

The pion multiplicities per event at 400$A$ MeV are 
$<n_{\pi^{-}}> = 0.92 \pm 0.17 (sys)$ and 
$<n_{\pi^{+}}> = 0.61 \pm 0.11 (sys)$
for the most central 10 \% of $\sigma_{geom}$.
The resulting $<n_{\pi^{-}}>/<n_{\pi^{+}}> = 1.51 \pm 0.39 (sys)$
agrees with the isobar model calculation \cite{rst1}, which is shown 
by the solid line in the bottom panel of Fig.~\ref{Fig:dndy4}, 
within the quoted errors. Similarly, the pion multiplicities per event 
at 1528$A$ MeV are 
$<n_{\pi^{-}}> = 13.2 \pm 1.6 (sys)$,  
$<n_{\pi^{+}}> = 11.0 \pm 1.3 (sys)$, yielding
$<n_{\pi^{-}}>/<n_{\pi^{+}}> = 1.2 \pm 0.2 (sys)$,
shown in the bottom panel of Fig.~\ref{Fig:dndy15}, 
again in nice agreement with the isobar model result.

\section{Thermal Model} \label{thermal}

In Sec.~\ref{spectra}, we have seen the indication that 
the rapidity distributions of pions in central Ru + Ru 
collisions are wider than the isotropic thermal source 
at midrapidity by comparing the data with the simplest 
thermal model formula, Eq.~(\ref{dndy_therm}) \cite{schne1}.
In order to investigate this observation further, 
a more instructive (but more complicated) thermal model 
is formulated in this section, including the decay of 
$\Delta(1232)$-resonances to pions explicitely.

The thermal model adopted in this paper has been described 
previously in Refs. \cite{wein1,wein2,hong4,hong2}. 
Decay pions from $\Delta(1232)$-resonances, $\pi_{\Delta}$, 
and additional thermal pions, $\pi_{T}$, 
are main ingredients in this model. For the momentum spectra 
of $\Delta(1232)$ and $\pi_{T}$ at freeze-out, 
we assume an isotropic expanding thermal source at the c.m. 
with the freeze-out temperature $T_{f} =$ 30 \cite{wre2} and 
84 MeV \cite{hong1} at 400$A$ and 1528$A$ MeV, respectively. 
Furthermore, the radial flow velocity $\beta_{f} =$ 0.3 is 
included for both beam energies. 
We use the following expression, which was proposed by 
Siemens and Rasmussen for the first time in late 70's \cite{rasm1},
for the thermal distributions of $\Delta(1232)$'s and $\pi_{T}$'s: 
\begin{eqnarray}
{{1} \over {2 \pi m_{t}}} {{d^{2}N} \over {dm_{t} dy^{(0)}}} &\propto&
E \cdot \exp \biggl( -{{\gamma E} \over  {T_{f}}}\biggr) \nonumber \\
 & & \times [(\gamma + {{T_{f}}\over{E}}){{\sinh \alpha} \over {\alpha}} -
{{T_{f}} \over {E}}\cosh \alpha],
\label{shellflow}
\end{eqnarray}
where $\gamma=1/\sqrt{1-\beta_{f}^{2}}$ and 
$\alpha=(\gamma \cdot \beta_{f} \cdot p)/T_{f}$.
Here $E = m_{t} \cosh y_{C}$ and 
$p = \sqrt{p_{t}^{2} + m_{t}^{2} \sinh^{2} y_{C}}$ 
are the total energy and respective momentum of particle in c.m. 

\begin{figure}[t!]
\begin{center}
\includegraphics[width=8.5cm]{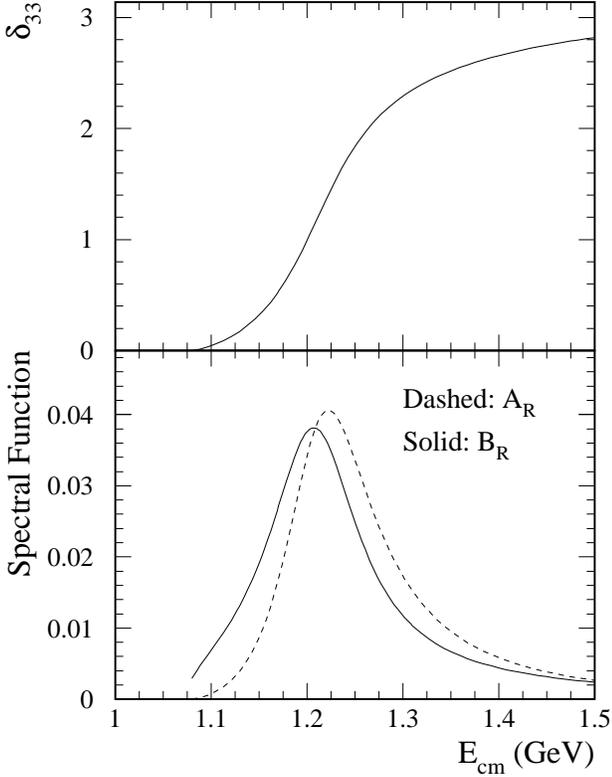}
\end{center}
\caption[]{
Phase shift $\delta_{33}$ (top) and spectral functions of the
$\Delta(1232)$-resonance (bottom). In the bottom panel, 
the dashed line represents a normal Breit-Wigner type 
function whereas the solid line represents 
the thermodynamic spectral function including 
contributions by the $\pi N$ interaction \cite{wein2}.}
\label{Fig:weinhold}
\end{figure}

Two different mass distributions of the $\Delta(1232)$ resonance 
have been considered with and without including the $\pi N$ 
interactions in the thermal fireball. Weinhold and collaborators have 
exploited the detailed calculation for the thermodynamic potential 
of a system consisting of pions and nucleons \cite{wein1,wein2}. 
Two $\Delta(1232)$-spectral functions are shown in the bottom 
panel of Fig.~\ref{Fig:weinhold}, where the dashed and solid lines 
are for the free and the modified spectral functions, respectively. 
The free spectral function $A_{R}$ is simply a normal Breit-Wigner 
shape 
\begin{equation}
A_{R}(E_{cm}) = {{\Gamma(E_{cm})} \over 
{(E_{cm}-E_{R})^{2} + {\Gamma(E_{cm})^{2}/4}}},
\label{Ar}
\end{equation}
where $\Gamma(E_{cm})$ is the energy-dependent 
width of the $\Delta(1232)$ \cite{koch1}:
\begin{equation}
\Gamma(E_{cm}) = 
\Gamma^{0}~\biggl( {{k_{0}} \over {k_{R}}}\biggr)^{3}~
{{E_{R}} \over {E_{cm}}}~ 
\biggl({{k_{R}^{2} + \delta^{2}} \over {k_{0}^{2} + 
\delta^{2}}}\biggr)^{2},
\label{gamma}
\end{equation}
where
\begin{equation}
E_{cm} = \sqrt{k_{0}^{2} + m_{N}^{2}} + 
\sqrt{k_{0}^{2} + m_{\pi}^{2}}
\label{kast}
\end{equation}
and
\begin{equation}
E_{R} = \sqrt{k_{R}^{2} + m_{N}^{2}} + 
\sqrt{k_{R}^{2} + m_{\pi}^{2}}
\label{kr}
\end{equation}
with $m_{N}$ being the nucleon mass.
The parameters used in this analysis are 
$\Gamma^{0} =$ 120 MeV, $E_{R} =$ 1232 MeV, 
and $\delta =$ 300 MeV, following Ref. \cite{wein2}. 

Now considering the $\pi N$ interactions, 
the spectral function is expected to be modified as
\begin{equation}
B_{R}(E_{cm}) = 2 {{\partial \delta_{33}(E_{cm})} 
\over {\partial E_{cm}}},
\label{Br}
\end{equation}
where the phase shift of the $\Delta(1232)$-resonance 
($P_{33}$-channel), $\delta_{33}$, can be deduced by
\begin{equation}
\tan \delta_{33}(E_{cm}) = - {{\Gamma(E_{cm})/2} 
\over {E_{cm}-E_{R}}}.
\label{d33}
\end{equation}
The phase shift factor $\delta_{33}$ is displayed in the upper 
panel, and the resulting $\Delta(1232)$-spectral function  
in the bottom panel of Fig.~\ref{Fig:weinhold}. 
It demonstrates a clear difference between the two spectral 
functions $A_{R}$ and $B_{R}$, especially close to the threshold; 
the modified function $B_{R}$ is shifted to lower masses. 
Note that $B_{R}$ can be uniquely determined by the measured 
phase shift $\delta_{33}$ from Eq.~(\ref{Br}) 
in a model-independent way. 

\begin{figure}[t!]
\begin{center}
\includegraphics[width=8.5cm,height=4.7cm]{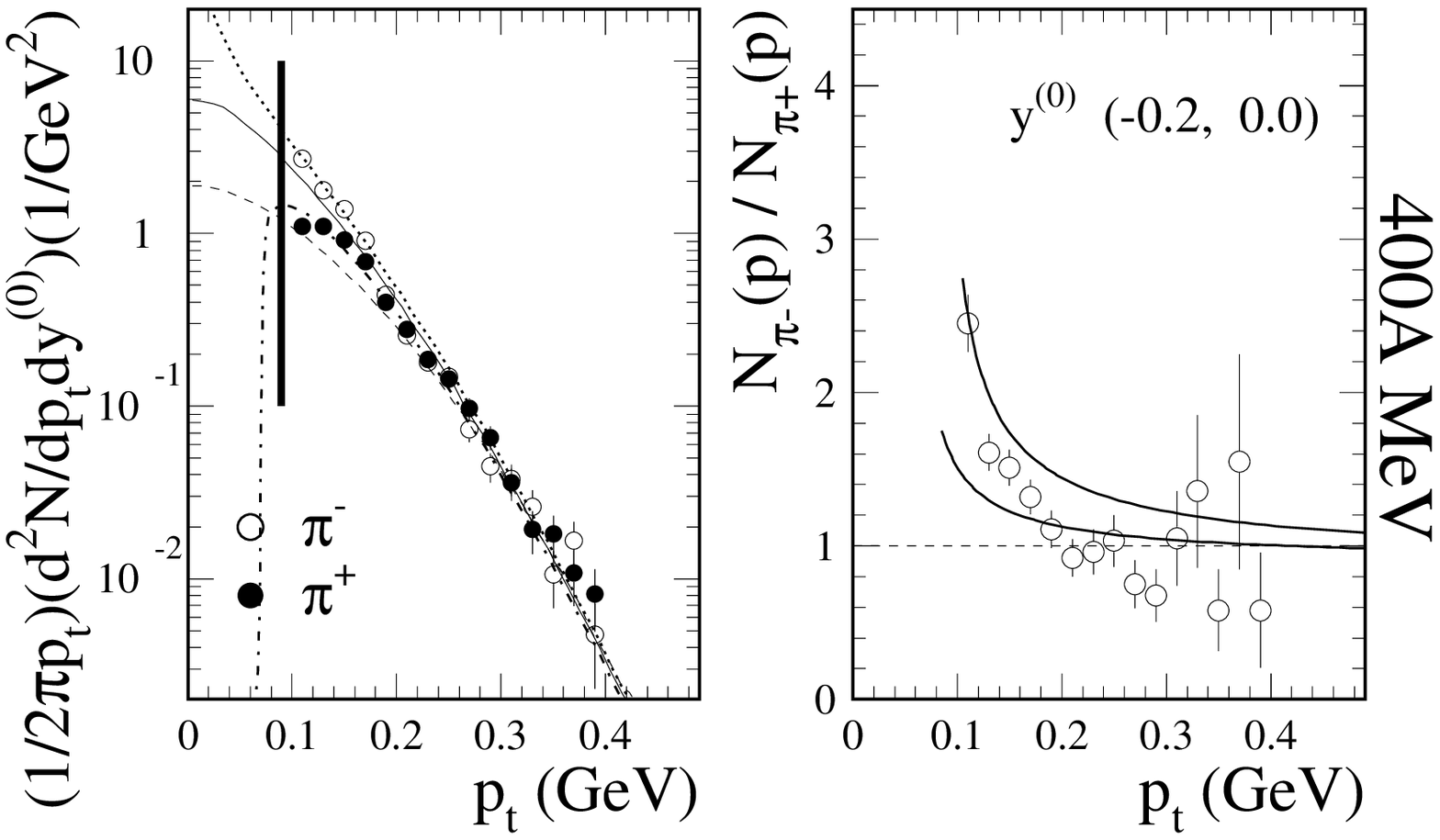}
\includegraphics[width=8.5cm,height=4.7cm]{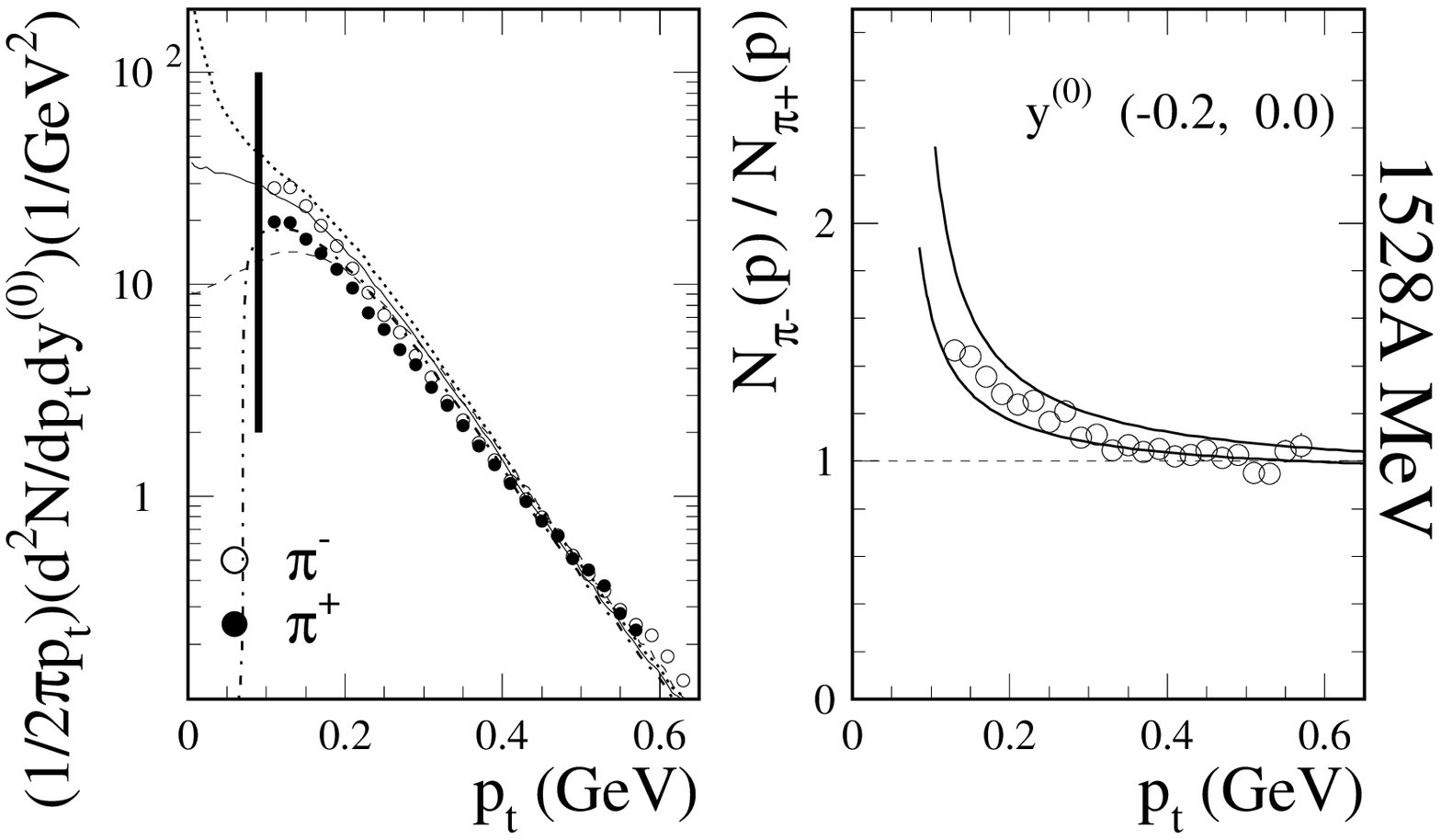}
\end{center}
\caption[]{ 
(Left) Invariant spectra of $\pi^{\pm}$ at midrapidity as a function 
of $p_{t}$ in comparison with the thermal model calculations 
at 400$A$ (top) and 1528$A$ MeV (bottom). 
The dashed line is obtained for the free $\Delta(1232)$-spectral function 
whereas the solid line is for the modified function (no Coulomb correction). 
The final results of the thermal model for $\pi^{-}$ and $\pi^{+}$ 
are shown by dotted and dash-dotted lines, respectively, 
after the static Coulomb potential effect is taken into account. 
Two thick vertical solid lines represent the low-$p_{t}$ limits for 
the validity of the present thermal model. 
(Right) Ratios of $\pi^{-}$ to $\pi^{+}$ yields at midrapidity. 
Two solid lines in each panel represent the upper and lower limits 
utilized in the present thermal model calculation.}
\label{Fig:therm}
\end{figure}

In the present thermal model, the $\Delta(1232)$-resonances 
in the thermally equilibrated system start to decay at the surface of 
the fireball at freeze-out. 
The resulting invariant spectra of $\pi^{\pm}$ near c.m. 
(now as a function of $p_{t}$) from the thermal model calculations 
are shown in the left panels of Fig.~\ref{Fig:therm}, 
where the dashed lines are obtained by $A_{R}$ and 
the solid lines are obtained by $B_{R}$. 
In the model calculations, thermal weighting factors are properly 
taken into account for both spectral functions.

\begin{figure}[t!]
\begin{center}
\includegraphics[width=8.5cm,height=4.5cm]{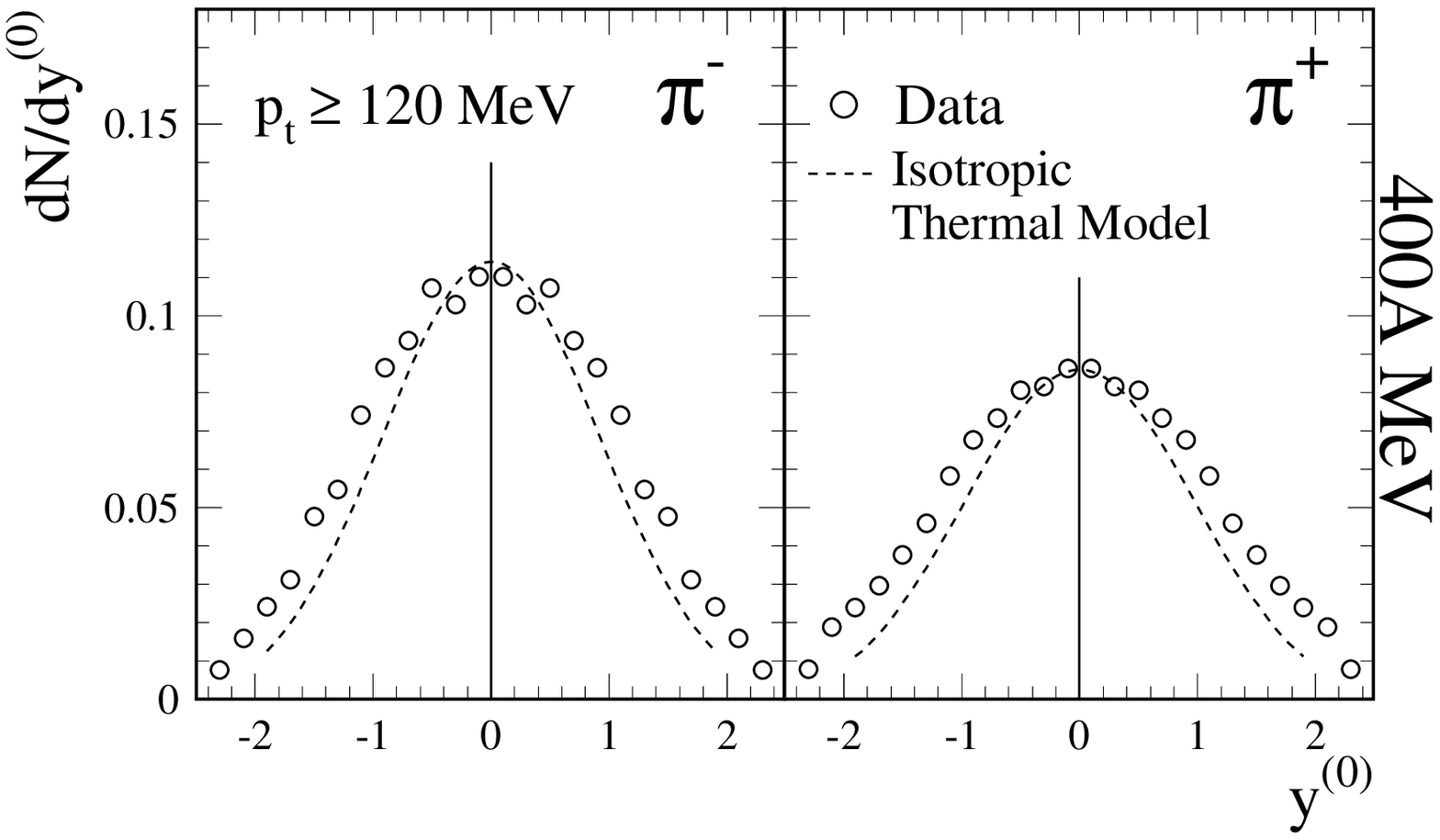}
\includegraphics[width=8.5cm,height=4.5cm]{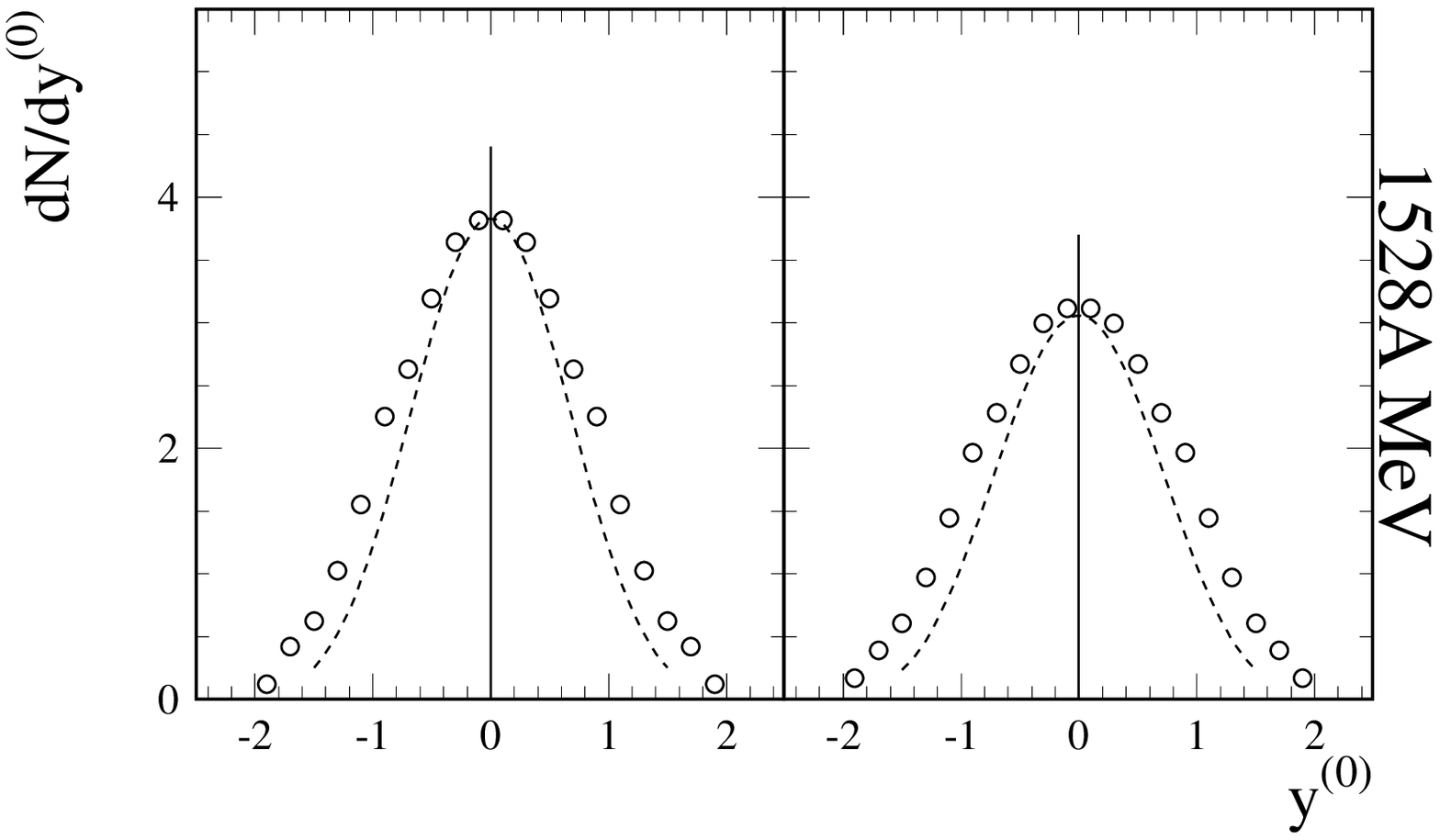}
\end{center}
\caption[]{ 
Measured rapidity spectra of $\pi^{-}$ (left) and $\pi^{+}$ (right) 
in comparison with the thermal model calculations at 400$A$ (top) 
and 1528$A$ MeV (bottom). 
The results from the thermal model, shown by the dashed lines, 
are obtained with isotropic thermal source at midrapidity.  
The modified $\Delta(1232)$-spectral function and 
Coulomb correction are included. 
Due to the low-$p_{t}$ limit of applicability for 
the present thermal model
(thick vertical solid lines in Fig.~\ref{Fig:therm}), 
the data and the model calculations are compared only 
for $p_{t} \geq$ 120 MeV.
}
\label{Fig:th_dndy}
\end{figure}

In Fig.~\ref{Fig:therm}, the ratios of pions from 
the $\Delta(1232)$-decay to all pions, 
$\pi_{\Delta}/(\pi_{\Delta}+\pi_{T})$, are determined by 
fitting the model calculations to the measured $p_{t}$ spectra; 
$\sim$ 72 \% at 400$A$ MeV and $\sim$ 76 \% at 1528$A$ MeV. 
Comparing to the previous results, the estimate at 1528$A$ MeV 
is about 12 \% higher than the number estimated for smaller
Ni + Ni collisions at similar beam energy (1450$A$ MeV) \cite{hong2}.
This implies that the fraction of projectile and target nucleons
excited to $\Delta(1232)$-resonances at freeze-out is 
higher for a larger collision system. But this ratio does not 
show the beam energy dependence between 400$A$ and 1528$A$ MeV.
One important aspect is that at 400$A$ MeV the ratio 
$\pi_{\Delta}/(\pi_{\Delta}+\pi_{T})$ is not uniquely determined 
with the transverse momentum spectra alone, as they can be 
well described by one exponential function. 
However, by fitting the transverse momentum spectra and 
the yields at midrapidity simultaneously, 
$\pi_{\Delta}/(\pi_{\Delta}+\pi_{T})$  can be uniquely fixed
(see Fig.~\ref{Fig:th_dndy} and the relevant text below).

From Fig.~\ref{Fig:therm} it is clear that the free spectral 
function $A_{R}$ can not reproduce data at low $p_{t}$ region 
($p_{t} \leq 250$ MeV), and that the modified spectral function 
$B_{R}$ can enhance the yields at low $p_{t}$, 
which brings the model calculation closer to the measurement. 
The same effect has also been observed before at similar beam 
energies between 1$A$ and 2$A$ GeV \cite{wein1,wein2,hong4,eskef}.

The contributions by higher baryonic resonances, 
e.g., $N^{\ast}$(1440), are negligible 
because of relatively low freeze-out temperature. 
Even in 1.9$A$ GeV Ni + Ni collisions the total contribution 
by higher mass resonances to total pion yield has been estimated 
to only $\sim$ 5 \% \cite{hong2}, hence this factor is neglected 
in the present thermal model analysis for the pion yields. 
Besides the yield estimation, higher mass resonances can not 
explain the differences in transverse momentum spectra 
between the data and the model calculations 
with $A_{R}$ in the left panels of Fig.~\ref{Fig:therm}. 
If we increase the number of higher mass resonances 
to compensate the difference at low $p_{t}$ 
by $N^{\ast} \to N + n \pi$ channels with $n \geq 2$, 
the pion contribution by $N^{\ast} \to N + \pi$ channels 
also increases accordingly because the branching ratios 
are fixed by the experimental data \cite{pdg}. 
But the decay momenta of pions in $N^{\ast} \to N + \pi$ 
processes are usually larger than the ones in 
the $\Delta(1232)$-decay so that the shape of the spectra 
at higher $p_{t}$ becomes much flatter than the data. 
Hence we conclude that the modification of the $\Delta$-spectral 
function is required to describe the present pion spectra.

As next step, we consider the effect of the Coulomb potential 
in order to explain the observed difference between 
the $\pi^{-}$ and $\pi^{+}$ spectra in the low-$p_{t}$ region. 
With a static approximation for the Coulomb field \cite{wagn1,e877-1,kapu1}, 
which neglects the time evolution of the fireball, $\pi^{\pm}$'s 
at freeze-out feel the effective Coulomb potential given by
\begin{equation}
V_{C} = {{Z_{eff} \cdot e^{2}} \over {R_{f}}},
\label{Vc}
\end{equation}
where $R_{f}$ is the radius of the fireball at freeze-out, 
$Z_{eff}$ the effective charge contained in the fireball. 
Then, the total energy of the emitted pion is modified, 
depending on their charge, in the following way:
\begin{equation}
E(p) = E(p_{i}) \pm V_{C},
\label{epi}
\end{equation}
where $p_{i}$ is the initial pion momentum without Coulomb potential. 
Now the number of particles with momentum $p$, 
$N(p) = d^{3}N/dp^{3}$, can be related to $N(p_{i})$ by the Jacobian:
\begin{equation}
N(p) = {\Bigg\vert}
{{\partial p_{i}^{3}} \over 
{\partial p^{3}}}{\Bigg\vert} N(p_{i})
= {{p_{i} E(p_{i})} \over {p E(p)}} N(p_{i})
= C^{\pm} N(p_{i})
\label{Np}
\end{equation}  
by using the identities $p \partial p = E \partial E$ 
and $\partial E(p) = \partial E(p_{i})$, where
\begin{eqnarray}
C^{\pm} 
&=& \sqrt{p^{2} \mp 2E(p)V_{C} + V_{C}^{2}} \cdot
{{E(p) \mp V_{C}} \over {p E(p)}} 
\label{Cpm}
\end{eqnarray}  
for positive ($C^{+}$) and negative ($C^{-}$) particles.
In order to determine the strength of the Coulomb potential, $V_{C}$, 
the experimental yield ratios of $\pi^{-}$ to $\pi^{+}$, i.e., 
$N_{\pi^{-}}(p) / N_{\pi^{+}}(p)$, 
at a given momentum $p$ are fitted by
\begin{eqnarray}
{{N_{\pi^{-}}(p)} \over {N_{\pi^{+}}(p)}} &=& 
{C_{R}}~{{\sqrt{p^{2} + 2E(p)V_{C} + V_{C}^{2}}} \over 
{\sqrt{p^{2} - 2E(p)V_{C} + V_{C}^{2}}}} \nonumber \\
&& \times \biggl( {{E(p) + V_{C}} \over {E(p) - V_{C}}} \biggr)~
{{N_{\pi^{-}}(p_{i})} \over {N_{\pi^{+}}(p_{i})}},
\label{ratio}
\end{eqnarray}
where the normalization constant $C_{R}$ and 
the Coulomb potential $V_{C}$ are two free fit 
parameters to be determined by the data. 
The constant $C_{R}$ is responsible for the height 
whereas $V_{C}$ determines the slope of 
$N_{\pi^{-}}(p) / N_{\pi^{+}}(p)$ at low $p_{t}$. 
The right panels of Fig.~\ref{Fig:therm} show 
the comparisons of the data with thermal model calculations 
at both beam energies, assuming the same momentum dependence 
of $\pi^{\pm}$ spectra before the Coulomb correction, 
$N_{\pi^{-}}(p_{i}) = N_{\pi^{+}}(p_{i})$ in this case. 
The values of $C_{R} = 0.9$ and $V_{C} = 17 \pm 6$ MeV 
describe data well at 400$A$ MeV, especially in the low-$p_{t}$ 
region where the $\pi^{\pm}$ spectra differ significantly. 
Similarly, $C_{R} = 0.9$ and $V_{C} = 16 \pm 4$ MeV are 
the best set of parameters at 1528$A$ MeV. 
The agreement between the data and the model calculations 
in $N_{\pi^{-}}(p) / N_{\pi^{+}}(p)$
at midrapidity is reasonable at both beam energies. 
Note that the same analysis technique as in this paper 
can reproduce the $\pi^{\pm}$ spectra, measured in 
Au + Au collisions at 1$A$ GeV by the KaoS collaboration, 
with $V_{C} =$ 25 MeV \cite{hong4}. 

It should be emphasized that the dynamical consideration 
becomes important only when the velocity of pions 
is smaller than the surface expansion velocity ($\beta_{s}$) of 
the fireball \cite{kapu1}. Charged pions with the velocity 
larger than $\beta_{s}$ see a time-independent Coulomb potential, 
and the present thermal model is applicable only 
for those pions.
For the average radial flow velocity $\beta_{f} = 0.3$, 
$\beta_{s}$ is about 0.40 (or 0.53) by assuming a linear 
(or quadratic) flow profile as a function of the fireball radius. 
Then, the critical momentum ($m_{\pi} \gamma_{s} \beta_{s}$) of 
pion is about 60 (90) MeV for the linear (quadratic) flow profile.
These estimated limits, shown by thick vertical 
solid lines in the left panels of Fig.~\ref{Fig:therm}, 
are close to or slightly lower than the low-$p_{t}$ 
limit of our measured pion spectra, which is about 100 MeV.

By using the determined values of $V_{C}$, 
we can generate the invariant spectra of $\pi^{\pm}$ 
by Eqs.~(\ref{Np}) and (\ref{Cpm}) at 400$A$ and 1528$A$ MeV. 
The dotted and dash-dotted lines in the left panels 
of Fig.~\ref{Fig:therm} represent the final invariant spectra 
obtained by the thermal model for $\pi^{-}$ and $\pi^{+}$, 
respectively, after the Coulomb correction is included. 
Good agreement between the data and the thermal model 
calculations can be found.

Once having estimated the magnitude of the Coulomb potential, 
one can deduce the source size at thermal freeze-out by using 
Eq.~(\ref{Vc}). The centrality condition used in this analysis 
corresponds to the average geometrical impact parameter 
$<b_{geom}>$ of $\sim$ 2.3 fm. 
The evaluated number of participant nucleons ($A_{part}$) can be 
estimated to about 146 by employing the recipe given in Ref. \cite{gos1}. 
After scaling with $Z/A$ of projectile and target, one obtains 
$Z_{eff} \approx 66.7$. The corresponding freeze-out radii  
$R_{f} = Z_{eff} \cdot e^{2} / V_{C}$ are 5.6 $\pm$ 2.0 fm 
at 400$A$ MeV and 6.0 $\pm$ 1.5 fm at 1528$A$ MeV. 
These results demonstrate that the thermal freeze-out 
radius is almost independent of the beam energy.
We also compare the freeze-out radii estimated in this paper
with two other published values that were obtained 
by using the same method. Having applied the present 
thermal model to the KaoS spectra at the SIS, 
the pion freeze-out radius was estimated as 6.3 $\pm$ 0.5 fm 
for Au + Au collisions at 1$A$ GeV \cite{hong4}.
When the same model was applied to the E877 spectra at the AGS, 
the pion freeze-out radius was estimated as 6.4 $\pm$ 4.5 fm 
for Au + Au collisions at 10.8$A$ GeV \cite{e877-1}. 
Although the error in the AGS result is too big to draw 
any meaningful conclusion, the mean radii do not show 
any beam energy dependence for Au + Au collisions, too. 
It is not possible to compare directly the present results 
with other numbers mentioned above because 
the system size is not the same 
(the mass number of Au is more than twice that of Ru).
Therefore, we compare the ratio of the estimated freeze-out 
radius $R_{f}$ to the radius of the projectile (or target) nucleus.
The results are summarized in Table~\ref{Tab:radius}; 
the resulting ratios are very similar 
at different beam energies between 0.4$A$ and 10.8$A$ GeV 
despite of a small difference in the collision centrality.

\begin{table}[t!]
\caption{Comparison of the ratio of the estimated pion 
freeze-out radius $R_{f}$ to the sharp-sphere radius $R_{0}$ 
(= 1.2$A^{1/3}$ fm) of the projectile (or target) nucleus 
at different beam energies ($E_{b}$).}
\begin{ruledtabular}
\begin{tabular}{cccc}
$E_{b}$ ($A$ GeV) & System & Centrality (\%) & $R_{f}/R_{0}$ \\ \hline
0.4 & Ru + Ru & 10 & 1.0 $\pm$ 0.4 \\
1.0 & Au + Au & 14 & 0.9 $\pm$ 0.2 \\
1.5 & Ru + Ru & 10 & 1.1 $\pm$ 0.3 \\
10.8 & Au + Au & 4 & 0.9 $\pm$ 0.6\\
\end{tabular}\label{Tab:radius}
\end{ruledtabular}
\end{table}

Only the midrapidity $p_{t}$ spectra of pions are tested for 
the present thermal model .
At forward and backward rapidities, the situation is more 
complicated because of the observed incomplete stopping 
and partial transparency of projectile and target nuclei 
at these beam energies \cite{hong1,hong3,hong_panic,willy04}. 
Figure~\ref{Fig:th_dndy} shows the measured rapidity spectra 
of $\pi^{\pm}$'s in comparison with the thermal model calculations;
the data and the model calculations are compared only 
for $p_{t} \geq$ 120 MeV due to the uncertainty of the spectral shape 
at the low-$p_{t}$ region in the present thermal model, which are 
indicated by thick vertical solid lines in Fig.~\ref{Fig:therm}. 
This comparison demonstrates that the thermal model calculations, 
assuming isotropic midrapidity thermal source, give somewhat 
narrower rapidity distributions than the data at both beam energies.
In order to reproduce the full measured rapidity spectra, 
the longitudinally elongated, rather than isotropic, thermal source 
of $\Delta(1232)$-resonances and pions should be assumed.
Although this kind of adjustment is, in principle, possible by changing 
the widths of the rapidity distributions of $\Delta(1232)$'s 
and $\pi_{T}$'s, we do not attempt to calculate 
the exact rapidity spectra of pions at present.

\section{Comparison with IQMD} \label{iqmd}

\begin{figure}[t!]
\begin{center}
\includegraphics[width=8.5cm]{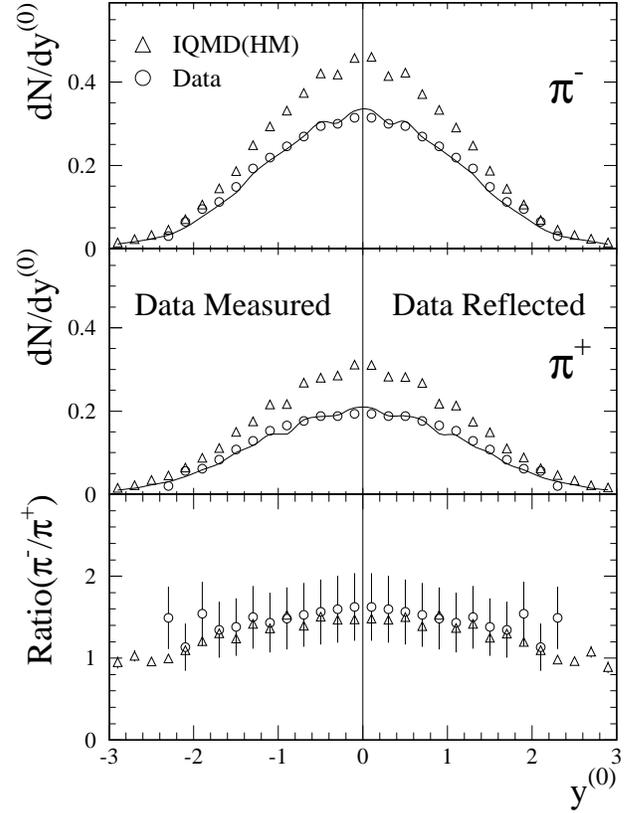}
\end{center}
\caption[]{
Comparison of data (circles) with the IQMD(HM) calculations 
(triangles) for the rapidity distributions of $\pi^{-}$ (top) and 
$\pi^{+}$ (middle) in Ru + Ru collisions for the most central 
10 \% of $\sigma_{geom}$. 
Solid lines in these two panels show the IQMD(HM) calculations 
normalized to the total experimental yields of $\pi^{\pm}$. 
The bottom panel shows the ratio of the $\pi^{-}$ and 
$\pi^{+}$ distributions, both for data and model results. 
Results from IQMD(SM) are almost identical to those from 
IQMD(HM) so that we omit in these plots.}
\label{Fig:dndy_iqmd4}
\end{figure}

\begin{figure}[t!]
\begin{center}
\includegraphics[width=8.5cm]{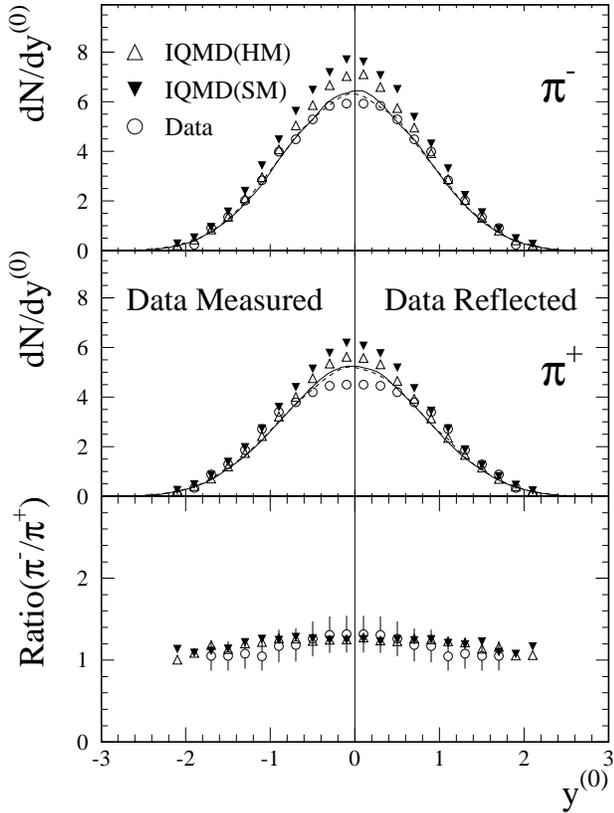}
\end{center}
\caption[]{
Same as Fig.~\ref{Fig:dndy_iqmd4}, but for 1528$A$ MeV.
Inversed filled triangles represent the IQMD(SM) calculation.
Dashed lines in top two panels show the IQMD(SM) calculations 
normalized in such a way that total yields of $\pi^{\pm}$ 
are the same as data.}
\label{Fig:dndy_iqmd15}
\end{figure}

\begin{figure}[t!]
\begin{center}
\includegraphics[width=8.5cm,height=4.7cm]{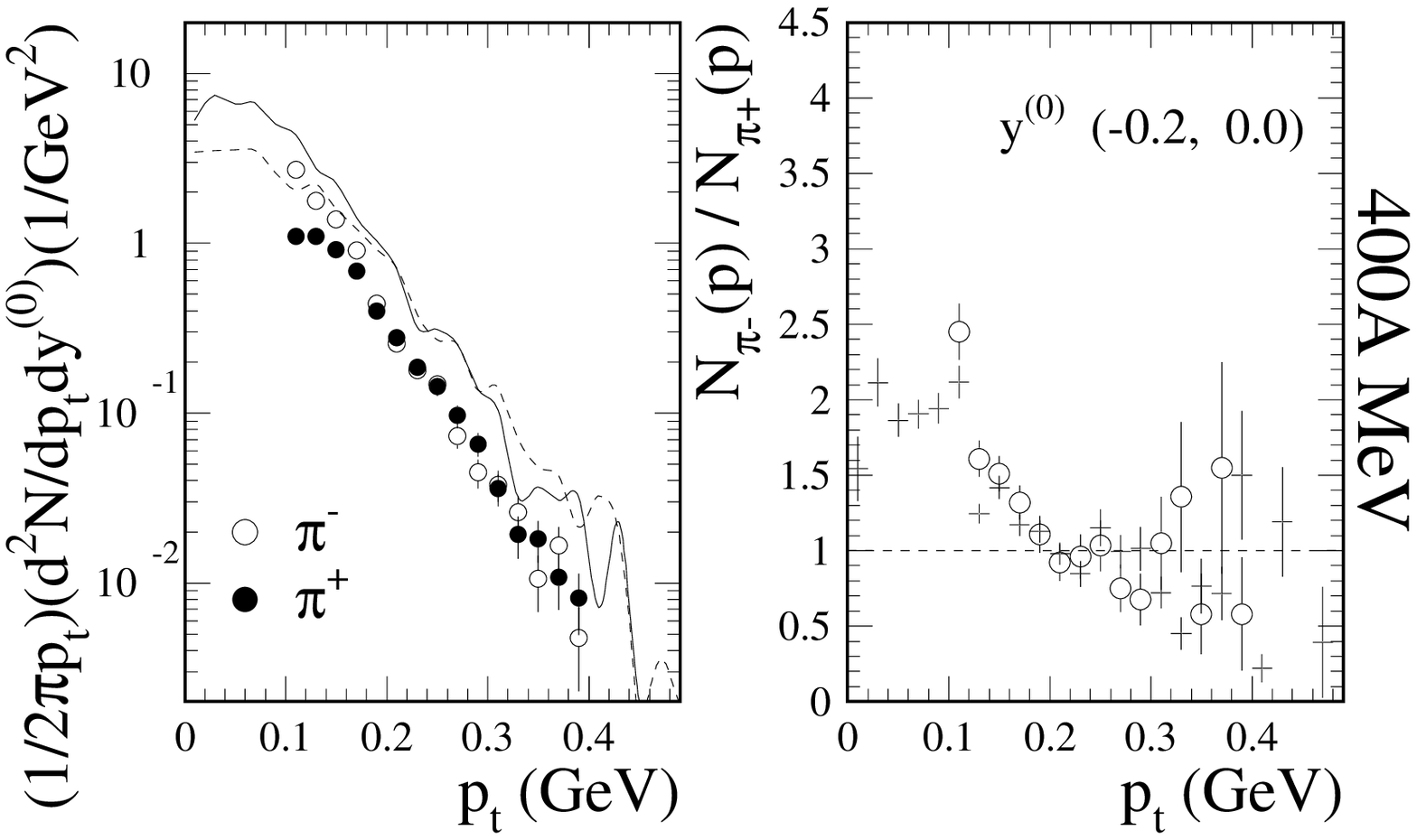}
\includegraphics[width=8.5cm,height=4.7cm]{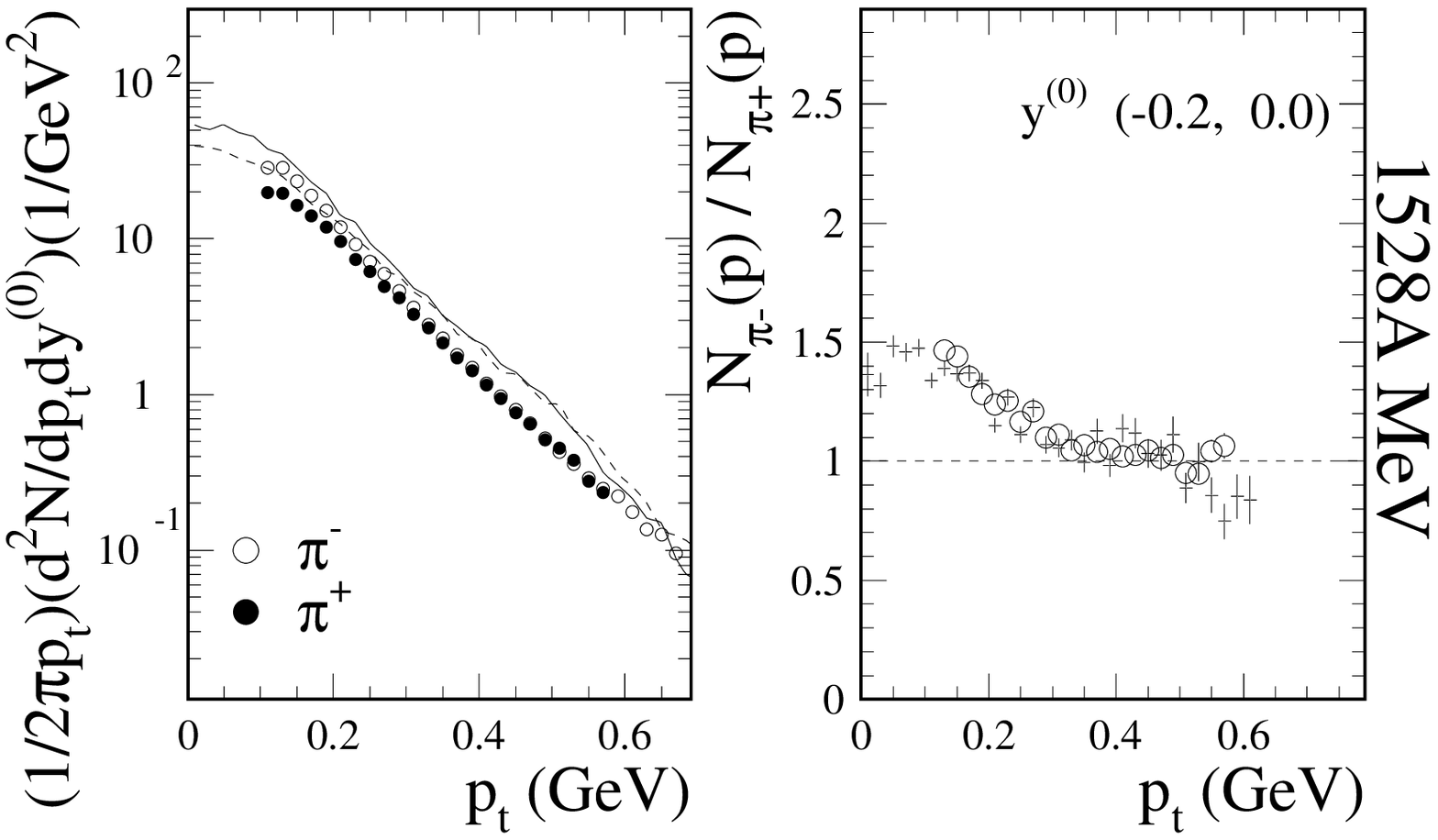}
\end{center}
\caption[]{
(Left) Invariant spectra of $\pi^{\pm}$ at midrapidity 
as a function of $p_{t}$ in comparison with the IQMD(HM) 
calculations, shown by solid and dashed lines 
for $\pi^{-}$ and $\pi^{+}$, respectively.
(Right) Ratio of $\pi^{-}$ to $\pi^{+}$ yields at midrapidity. 
Crosses are from the IQMD(HM) calculations.
Results from the IQMD(SM) are very similar to those from
the IQMD(HM) except about 10 \% higher yield at 1528$A$ MeV.
Top and bottom panels are for the data and the model 
calculations at 400$A$ and 1528$A$ MeV, respectively.}
\label{Fig:pt_iqmd}
\end{figure}

The Isospin Quantum Molecular Dynamics (IQMD) model 
is a nonequilibrium transport model which considers 
the isospin degree of freedom for the nucleon-nucleon 
cross section and the Coulomb interaction \cite{hart1}. 
In the framework of the IQMD model, pions are explicitly 
formed via the decay of the $\Delta$-resonances,
and experimental cross sections are considered.
The IQMD version utilized in this analysis contains only 
the lowest three $\Delta$-states. 
A modified detailed balance treatment, which accounts 
for their finite width, is also considered. 
As a consequence, the appearing spectral shape of $\Delta$'s 
is changing in the course of the reaction.
Therefore, in the charged pion spectra, the IQMD model has to 
show, in principle, the effects of the Coulomb interaction 
in addition to other collision effects 
including the $\pi N$ interactions. 

IQMD has been successful in reproducing various aspects 
of heavy ion collisions, including the pion production 
in Au + Au collisions at 1$A$ GeV \cite{bass1}. 
In this section, we compare the present experimental pion spectra 
with the IQMD calculations in Ru + Ru collisions at 400$A$ and 1528$A$ MeV. 
For these calculations, a hard (compressibility $K$ = 380 MeV)
and a soft ($K$ = 200 MeV) EoS versions are used,
including the momentum dependence of the nucleon interaction (MDI); 
they are denoted by IQMD(HM) and respective IQMD(SM) in the following.
Note that the IQMD events are analyzed with the same analysis 
procedure as the data. As in the data analysis, we also use 
the $E_{rat}$ distribution at 400$A$ MeV to select the most central 
10 \% of $\sigma_{geom}$ whereas at 1528$A$ MeV 
the total multiplicity measured in the PLAWA and 
the CDC acceptance is taken. 

Top and middle panels of Figs.~\ref{Fig:dndy_iqmd4} 
and ~\ref{Fig:dndy_iqmd15} show the comparisons 
of the experimental $dN/dy^{(0)}$ spectra (circles) 
with the IQMD(HM) calculations (open triangles) 
for $\pi^{-}$ and $\pi^{+}$, respectively. 
The IQMD(SM) calculations (inverted solid triangle) are almost 
identical with IQMD(HM) at 400$A$ MeV, but they are about 
10 \% larger in yields for both charges at 1528$A$ MeV.
Therefore, we include these results only in 
Fig.~\ref{Fig:dndy_iqmd15} at the higher beam energy.

In accordance with earlier investigations \cite{marc1}, 
we find that the IQMD model produces too many pions 
in Ru + Ru collisions, especially at 400$A$ MeV.
In order to compare the shape of the $dN/dy^{(0)}$ spectra, 
we need to scale the IQMD results by 0.72 and 0.67 
for $\pi^{-}$ and $\pi^{+}$, respectively, 
at 400$A$ MeV for both HM and SM
(solid lines in Fig.~\ref{Fig:dndy_iqmd4}).
These scale factors are determined in such a way 
that the integration of the model $dN/dy^{(0)}$ 
yields the same value as the integrated data distribution. 
Similarly, we also need to scale the IQMD results by 
0.90 for $\pi^{-}$ and 0.93 for $\pi^{+}$ at 1528$A$ MeV 
with HM (solid lines in Fig.~\ref{Fig:dndy_iqmd15}).
On the other hand, this scaling factor is reduced to 
0.82 for $\pi^{-}$ and 0.84 for $\pi^{+}$ with the choice 
of SM (dashed lines in Fig.~\ref{Fig:dndy_iqmd15}).
The scaled $dN/dy^{(0)}$ distributions from 
the IQMD calculations can nicely reproduce 
the shape of the measured $dN/dy^{(0)}$ spectra of pions 
at both beam energies. 
The bottom panels of Figs.~\ref{Fig:dndy_iqmd4} 
and ~\ref{Fig:dndy_iqmd15} show that the 
ratios of the $dN/dy^{(0)}$ distributions 
of $\pi^{-}$ and $\pi^{+}$ agree with the data 
within errors at both beam energies.

The left panels of Fig.~\ref{Fig:pt_iqmd} display 
the comparisons of data with the model calculations 
for the invariant spectra of $\pi^{\pm}$ at midrapidity. 
Similar to the experimental data, the IQMD(HM) 
calculations show different shapes for $\pi^{-}$ 
and $\pi^{+}$ in the low transverse momentum region, 
which can be attributed to the Coulomb interaction. 
These differences can be presented more clearly 
in the ratio $N_{\pi^{-}}(p)/N_{\pi^{+}}(p)$ as 
a function of $p_{t}$ (right panels of Fig.~\ref{Fig:pt_iqmd}).
The IQMD(HM) calculation agrees very nicely with the data.
The IQMD(SM) calculation gives very similar results 
on the transverse momentum spectra of pions and 
their ratios except about 10 \% higher production
than HM at 1528$A$ MeV.

\section{conclusions}\label{conc}

We have presented charged pion spectra in central 
Ru + Ru collisions at 400$A$ and 1528$A$ MeV. 
In both beam energies the uppermost 10 \% of the geometric 
cross section have been selected. 
The measured transverse momentum spectra 
in their invariant form as a function of $m_{t}$ can be described 
very well by one exponential function at 400$A$ MeV and 
by a sum of two exponential functions at 1528$A$ MeV. 

The results from thermal model have been compared to 
the transverse momentum spectra of pions at midrapidity. 
Pions from the decay of the thermal $\Delta(1232)$-resonances 
in addition to thermal pions are main ingredients of 
the present thermal model.
The momentum distributions of $\Delta(1232)$-resonances
and thermal pions at freeze-out are assumed to be an expanding 
isotropic source at the center of mass. 
Pion spectra at low transverse momentum require the modification 
of the $\Delta(1232)$-spectral function due to the $\pi N$ 
interactions at both beam energies. 
The Coulomb potential is necessary to explain 
the different spectral shapes between the $\pi^{-}$ and $\pi^{+}$ 
spectra in the low transverse momentum region. 
With this we find excellent agreement between the data 
and the thermal model calculations. 
Using the estimated strength of 
the Coulomb potential, which is determined by 
the momentum dependent yield ratio of $\pi^{-}$ to $\pi^{+}$, 
the freeze-out radius of the fireball has been deduced 
to be about 6.0 fm at both 400$A$ and 1528$A$ MeV.
The rapidity spectra of pions require the longitudinally elongated 
(not isotropic) thermal source of $\Delta$'s and pions.

Finally, the experimental data are compared to 
IQMD model calculations. Although the absolute yields 
of charged pions in the model are somewhat larger
(about 30 and 10 \% at 400$A$ and 1528$A$ MeV, respectively), 
the shapes of the rapidity and transverse momentum spectra 
agree nicely with the data. 
The IQMD model displays that the yield ratios 
of $\pi^{-}$ to $\pi^{+}$ at low transverse momenta 
deviate from unity due to the Coulomb interaction; 
the agreement between the data and the IQMD model is satisfactory.
The pion yield is independent of the choice of the equation of state 
at 400$A$ MeV, whereas it is about 10 \% higher with a soft than 
with a hard EoS at 1528$A$ MeV . 
The comparison of the pion transverse momentum and rapidity spectra 
with two completely different approaches (thermal model and IQMD) 
renders us the importance of 
the Coulomb interaction and the collision effects including 
the $\pi N$ interactions in the pion spectra at SIS energies.

\begin{acknowledgments}

We gratefully acknowledge support from 
the Korea Research Foundation (KRF) under 
Grant No. KRF-2002-015-CS0009, and from the 
Deutsche Forschungsgemeinschaft (DFG)
under the project No. 446 KOR-113/76/0. 
\end{acknowledgments}

\end{document}